# Steps and bumps: precision extraction of discrete states of molecular machines using physically-based, high-throughput time series analysis

Max A. Little[1,2], Bradley C. Steel[2], Fan Bai[3], Yoshiyuki Sowa[4], Thomas Bilyard[5], David M. Mueller[6], Richard M. Berry[2], Nick S. Jones[1,2]

[1]Oxford Centre for Integrative Systems Biology, University of Oxford, UK, [2]Department of Physics, University of Oxford, UK, [3]Graduate School of Frontier Biosciences, Osaka University, Japan, [4]Department of Frontier Bioscience, Hosei University, Japan, [5]Lawrence Berkeley National Lab, Berkeley, CA, USA, [6]Department of Biochemistry and Molecular Biology, Rosalind Franklin University of Medicine and Science, The Chicago Medical School, North Chicago, IL, USA. Correspondence should be addressed to N.S.J. (n.jones1@physics.ox.ac.uk).

## Abstract

**We report new statistical time-series analysis tools providing significant improvements in the rapid, precision extraction of discrete state dynamics from large databases of experimental observations of molecular machines. By building physical knowledge and statistical innovations into analysis tools, we demonstrate new techniques for recovering discrete state transitions buried in highly correlated molecular noise. We demonstrate the effectiveness of our approach on simulated and real examples of step-like rotation of the bacterial flagellar motor and the $F_1$-ATPase enzyme. We show that our method can clearly identify molecular steps, symmetries and cascaded processes that are too weak for existing algorithms to detect, and can do so much faster than existing algorithms. Our techniques represent a major advance in the drive towards automated, precision, high-throughput studies of molecular machine dynamics. Modular, open-source software that implements these techniques is provided.**

## Introduction

Nature has evolved many molecular machines such as pumps, copiers and motors. Biophysical theory proposes that these machines, that convert electrochemical energy to linear or rotary motion, do so in a series of thermally driven, nanoscale "step-like" motions because this maximizes the use of the available free energy[1]. Even in genetically identical cells, each cell shows fundamental variability partly traced to thermal randomness in discrete molecular mechanochemistry[2]. In gene expression, intrinsic noise in the machinery of transcription affects protein copy numbers, and the phenotype can be highly sensitive to the number of proteins available to reactions. Machines such as kinesin moving along microtubules[3], myosin sliding between actin filaments[4], and rotations of protein complexes in the flagellar motor[5], all show the hallmarks of discreteness with superimposed thermal fluctuations. Motion is often highly repetitive and *(quasi)-periodic*, because these machines are built of many identical copies of molecular components coupled together in linear, circular, or helical patterns. Of interest to studies of these molecular machines are temporal sequences of discrete states observed using advanced experimental techniques such as Förster resonance energy transfer (FRET), back focal plane interferometry or atomic force microscopy[6]. Due to discreteness of step-like transitions, the distribution of states is often multimodal ("bump-like"). A common view of molecular machines is that they execute *randomly forced motion* in a *potential energy well* around each discrete state (described as *generalized Langevin dynamics*), leading to temporally correlated noise[1].

The dominant approach to extracting discrete states of molecular machines is to first smooth away (*filter*) as much noise as possible, leaving the underlying transitions[4,5,7,8]. Classical *running mean* filtering is fast and simple, but fundamentally inadequate because (a) it cannot separate steps from noise, (b) it is statistically optimal only if the noise is uncorrelated[9], (c) it operates on a fixed-length sliding time window of the series that favours certain durations between transitions, and (d) the observed stepping may not be instantaneous due to elastic coupling or friction effects[10]. Adaptations have been proposed[11-13], but none address all the above issues simultaneously. Sophisticated alternative algorithms (e.g. based on Markov Chain Monte Carlo, particle filtering or variational Bayesian techniques) might be able to tackle several of these problems. However, such algorithms would be too computationally demanding to rapidly process large numbers of time series, and both in principle and in practice, become intractable as the data size increases[14]. After filtering, finding the arrangement of the discrete states is a *bump-hunting* problem: finding peaks in a distribution. Histograms of the smoothed time series are easily constructed and popular[5,8], but the choice of bin edges and widths is an open-ended problem not solvable without making extra assumptions that may not be appropriate. Furthermore, the resulting histogram distribution estimate is discontinuous[15] and therefore an unrealistic representation of the underlying potential well(s)[1]. Many more sophisticated algorithms exist (such as *mixture modelling*[16]), but for the multiple states in typical molecular machines and high-throughput settings, these algorithms are currently computationally intractable. *Kernel methods* are continuous and more tractable, but kernel bandwidth and shape selection are open problems as with histogram parameters.

How can step analysis be carried out for time traces of (quasi)-periodic molecular dynamics corrupted by correlated and other mechanochemical noise, and how can this be achieved on realistic time scales for large data sets? Figure 1 illustrates our overall analysis process. This begins with time-based observations of the motion of the molecular machine. Typical experiments include optical trap or FRET measurements[6] of RNA polymerase transcription dynamics (time series of transcript length); position detection of laser-illuminated beads attached to linear and rotary motors[3,5], or blockade currents of hemolysin nanopores with translocating DNA[17]. We propose novel algorithms that handle correlated noise and non-instantaneous stepping by incorporating a simple physical model *directly* into the filter structure, in such a way that does not require the stipulation of an artificial window size, and that guarantees rapid discovery of the optimal solution without wasted computation. We then make use of contemporary statistical methods that permit us to find quasi-periodic arrangements of bumps in the distribution of molecular states by simple Fourier coefficient selection, thereby circumventing the need to estimate the distribution directly. Finally, we recover the discrete states by statistically optimal classification of the smoothed time trace to the nearest peak of each bump in the distribution.

Note that these techniques are equally applicable to both linear and rotary machines. To ensure that sufficient samples of the dwell states are represented, it is, however, sensible to restrict linear motion to a bounded interval. This is a very common physical situation – many linear molecular machines undergo repetitive sequences of state transitions. If they do not, then the motion can be wrapped onto a bounded interval whose length is a multiple of any fundamental state transition size before application of these new techniques, and the process is fundamentally the same.

We demonstrate by simulation that these new algorithms are a substantial improvement in accuracy, precision and speed over what is currently possible. Moving to real data, we then rapidly process a large number of experimental angle-time traces of the bacterial flagellar motor, and unambiguously identify symmetries in the discrete state locations. We are also able to find a very large number of dwell times to provide clear evidence for non-Poisson stepping in the flagellar motor, and resolve characteristic patterns of cascaded rate-limiting reactions in $F_1$-ATPase stepping. To accompany this paper, we have released modular, open source code that is freely available for download. This code implements the algorithms described in the main text and Supplementary Methods, and additional useful utility functions.

## Results

### Simulated data performance comparisons

Figure 2 details the step-smoothing and bump-hunting methods. Validation of our techniques is carried out on simulated bacterial flagellar motor rotation (Langevin dynamics, see Supplementary Methods) with known discrete states and state dynamics, over nine test cases that explore the variability of real biological recordings (see Figure 3 and caption for description of the test case parameters). We compare the step-smoothers in this paper (algorithms L1-PWC and L1-PWC-AR1, see Supplementary Methods) against the classical median filter[9], the Chung-Kennedy filter[11], and the Kalafut-Visscher step-finder[18] in terms of both the absolute state location recovery error (MAE) and relative absolute roughness (RAR) of the estimated time series (see Figure 3 and Supplementary Methods for detailed descriptions). The parameters for these step-smoothers are optimized for the best performance on these simulations (see Supplementary Methods). The average execution time for the algorithms was, in order of decreasing speed: median filter, five seconds; L1-PWC and L1-PWC-AR1, seven seconds; Chung-Kennedy filter 41 seconds, and Kalafut-Visscher algorithm 1,020 seconds (17 minutes).

The new step-smoothers clearly outperform existing methods. By design, median and Chung-Kennedy filters cannot guarantee that the recovered time trace of states will be constant when the machine is actually stationary, so both the recovery error and relative absolute roughness are worse than the L1-PWC filters. Similarly, the Kalafut-Visscher step-finder is confounded by the correlation in the noise, and so finds excessive detail, thus, the relative absolute roughness and the recovery error are large. We note that simulations with back-stepping lead to similar results.

We tested how well the thresholded empirical characteristic function (algorithm ECF-Bump, see Supplementary Methods) can recover the known, dominant periodicity of a simulated bacterial flagellar motor, when compared to existing techniques (see Supplementary Tables 1 and 2). This new algorithm outperforms the alternative methods that are based upon first estimating the distribution of discrete states. This comparison highlights some of the shortcomings of other plausible bump-hunting techniques. For histogram-based methods[5], the histogram bin width sets a fundamental limit on the maximum periodicity that can be identified: increasing the bin width decreases this frequency. However, reducing the bin sizes increases the error of the bin counts due to finite sample size effects. This bin count error is particularly problematic for high numbers of discrete states or significant asymmetries in state locations where this method returns a wide spread of values. Similarly, for peak-finding in kernel density estimates, the choice of kernel width is a limiting factor: too large, and a small bump in the distribution will be merged into nearby bumps, too small and spurious bumps will appear. As with the histograms, there is no way to retain small bumps representing real discrete states, and at the same time remove spurious bumps that are due to experimental confounds or finite sample size effects. This is because these methods lack *global information* about the symmetries in the distribution, making them uncompetitive with the ECF-based algorithm. These sorts of issues with existing techniques confound straightforward bump-hunting in the distribution.

### Experimental data

Figure 4 illustrates the process of applying the novel step-smoothing and bump-hunting techniques to a 4.2s long, single experimental time-angle trace of a rotating *E. coli* flagellar motor with attached 200nm bead; the same data were originally published as Figure 2 in reference Sowa *et al.*[5]. We were able to process the entire trace consisting of 10,000 samples at 2.4kHz sampling rate, in less than 0.5s on a standard PC (note that this is faster than real time). Given the flagellar hook stiffness[19] and bead size (200nm), this sample rate is too slow so that all stepping appears instantaneous. Insignificant autocorrelation at positive time lags confirms this, implying that state transitions appear as effectively instantaneous and this suggests the L1-PWC algorithm that does not consider correlated noise (see Supplementary Methods). The results of the analysis confirm previous findings of 26 discrete states[5]. This single trace also shows some evidence for 2-, 11- and 17-fold symmetry. Existing techniques have yet to identify conclusively which of those symmetries is a genuine property of the motor.

Our methods are robust and fast enough that they can extract discrete states from every time-angle trace in the database, without the need for prior hand-editing. We applied the same process as in Figure 4 separately to all six traces obtained from the same motor. Figure 5a shows symmetry analysis averaged over these six traces. This provides clear evidence that the 26- and 11-fold symmetries are properties of the motor, whereas the other periodicities in Figure 4a are most likely artefacts due to finite sample size effects in this one trace. The increased precision of these novel techniques has therefore allowed us to show that the weak evidence for 11-fold symmetry in the original study (Figure 4b in Sowa et al[5]) is, most likely, a real feature of the motor.

Our methods allow us to capture ~6,000 dwell times over these six traces, making it possible to characterize, with high statistical power, the distribution of dwell times. In Figure 5(b-e) we fit four different distributions to the dwell times, including an exponential model. There is sufficient data to resolve the extremes of the distribution, clearly indicating that the simple exponential is not a good fit and that the extremes of the distribution have much higher probability than either exponential or gamma distributions would imply. Our model based-technique thus reveals new features of bacterial flagellar motor stepping.

Figure 6 shows the process applied to an experimental time-angle trace of a single rotating, surface immobilized yeast $F_1$-ATPase molecule with a 60nm gold bead attached at 30μM [ATP] (approximately equal to the Michaelis constant ~$K_m$). The 0.27s trace, recorded at 30kHz, has significant autocorrelation and is able to resolve non-instantaneous stepping. Therefore, we used the L1-PWC-AR1 algorithm with parameter $a_1$ set equal to the autocorrelation at a time lag of one sample (0.8, see Supplementary Methods for more details). This implies a relaxation to within a fraction $e^{-1}$ of the stationary dwell state of approximately 0.15ms. This algorithm was effective at recovering the non-instantaneous stepping. The pattern of dominant symmetries (Figure 6a,b) is consistent with the existence of six-step rotation in the pattern $120° \times n$ and $120° \times n + 30°$ where $n$ = 1, 2, 3, closely confirming the findings of previous studies on the slower thermophilic $F_1$ at similar [ATP][20].

Given that the molecule is expected to have almost perfect three-fold symmetry, it is most likely that the asymmetry of the distribution of in Figure 6b is due to experimental confounds such as the loose linkage of the bead to the molecule. This asymmetry manifests as some amplitude in the 1, 2, and 4-fold ECF components. This asymmetry would make it difficult to fit a single model to all of the 120° domains. However, the robustness of our techniques means that after the discrete state extraction process, a total of 502 dwell times and a median of 83.5 dwell times per state were obtained. This was sufficient data to fit *separate* distribution models to the dwell times for *each* of the six states.

Analysis of the quality of fit of various distribution models for each of the six states revealed that different models are appropriate for each of the six states. The best model we could find (in terms of *Bayesian Information Criterion* over all states, see Supplementary Methods) was the gamma distribution with scale $k$ = 2 for the three states located at $120° \times n$, and the exponential model for each of the three states at $120° \times n + 30°$. This model fitted better than a gamma model for every state (negative BIC difference 3.9), and markedly better than an exponential model for every state (negative BIC difference 49.8). We found that a double exponential model (see Supplementary Methods) fitted to the $120° \times n$ states had equal rate parameters, so that a gamma model with scale $k$ = 2 obtained an indistinguishable rate parameter. These findings are consistent with an interpretation that has two cascaded rate-limiting processes at the $120° \times n$ state, and one process at the $120° \times n + 30°$ state.

We find that if the autocorrelation and non-instantaneous stepping is ignored by using the L1-PWC algorithm that assumes, as with existing techniques, instantaneous stepping, then at low values of the regularization term, the algorithm is confounded by the relaxation effect and it returns many small, spurious steps that "interpolate" the rounded edges. At the other extreme of large regularization, smaller steps are missed altogether. We therefore cannot find one single, optimal value of the regularization term. This is because of the fundamental mismatch between the assumption of independence in the model, and the temporal dependence in the signal.

## Discussion

Our step-smoothing algorithms address the problem of analyzing and recovering discrete state transitions obscured by correlated noise from time series generated by Langevin molecular motion. We were able to process bacterial flagellar motor bead assay time-angle traces faster than real-time. Subsequent application of our distribution bump-hunting algorithm uncovers symmetries in bacterial flagellar motor traces that have hitherto been hinted at, but largely hidden due to the theoretical and practical shortcomings of existing algorithms. By using our discrete state distribution estimation algorithm, we are able to recover the time series of discrete state transitions, from which analysis of the distribution of dwell times shows significant departures from classical Poisson stepping (revealed by the divergence from exponential behaviour for extreme dwell times). To place these novel findings in context, a recent cryo-electron microscopy structure of the flagellar rotor indicates that there is a symmetry mismatch between different parts of the rotor, which varies from one motor to the next[21]. Non-Poisson stepping might be explained by static heterogeneity: mixed periodicities imply that steps need not be equivalent at all angles, but sufficiently large data-sets and robust analysis techniques such as presented here, may reveal simple Poisson stepping at each angle. Alternatively, heterogeneity may be dynamic, with the state of the motor changing in time due to exchange of stator complexes[22] or other regulatory processes. Analysis using our new methods of stepping traces from many motors will therefore be an important tool in the task of understanding heterogeneity in flagellar rotation and finding a model of the flagellar mechanism that explains the symmetries observed in both structural and rotation data.

The $F_1$-ATPase bead assay shows clear evidence of non-instantaneous stepping, with the relaxation time of the system about five times slower than the sampling duration. Therefore, by using Langevin dynamics in the L1-PWC-AR1 algorithm, we could extract symmetries that clearly revealed six states in one revolution of the enzyme in a characteristic angular pattern. We were able to extract a sufficient number of dwell times to detect differences in the number of rate-limiting processes responsible for each dwell. Current models of this rotary enzyme propose that each 120° step comprises an ADP release with ATP binding phase, followed by 80 – 90° rotation, an ATP cleave with Pi release phase, then a final 30 – 40° rotation; our findings are consistent with this interpretation[20]. Previous studies have not addressed the issue of which distribution best fits these dwell times, the analysis provided by this study therefore provides evidence supporting this model.

These findings with bacterial flagellar and $F_1$-ATPase motors are important because they build a direct connection between mathematical models of these systems, and high-throughput time series analysis of experimental observations. The major obstacles to exploiting modern high-throughput experimental data from molecular machines can be loosely categorized as *physical*, *statistical* and *practical*. The physical limitations are the unavoidable facts of life at the single-molecule level, such as shot and other instrumental noise, correlated Brownian noise and discreteness[1,6]. These can have a serious impact on accurate estimation of discrete state transitions and dwell times for example, and we demonstrated that increased precision is possible if these physical aspects are directly incorporated into the filtering of the time trace. Some physical hindrances can be turned into opportunities if they are handled appropriately. For example, the periodic arrangement of multiple molecules into complexes sets global constraints on the pattern of discrete states. This allowed us to make use of the powerful yet simple contemporary statistical theory of nonlinear threshold estimation to recover discrete state locations with high precision, because the symmetries of the molecular machine are naturally represented in the Fourier basis.

The predominant statistical limitations arise due to small sample sizes. In high-throughput experiments, this limitation is not due to a paucity of recorded data; rather, it stems from the limitations of the analysis tools. As a case in point, if non-instantaneous stepping forces the experimenter to discard much of the data because it is not of high enough "quality" to be analysed by a step-filtering algorithm, then there may not be enough stepping events left to obtain sufficient statistical power when comparing dwell time distribution models. This is a particular problem for rare events. These events occur, by definition, with low probability. Increasingly, it is being recognized that rare events are not irrelevant anomalies, but may play a central role in normal biological function. It is thus of importance to quantify their frequency, and we showed that we could achieve high exploitation rates of high-throughput data to get statistically robust estimates.

Practical barriers to analysis of high-throughput time series data are often computational or related to algorithmic complexity. Our algorithms make use of powerful yet simple contemporary statistical and computational methods. For example, least-squares, $L_1$-penalized estimation algorithms can be solved by simple, off-the-shelf convex optimization techniques, and have guarantees on recovery of the optimal solution that, for non-convex problems, could otherwise require expensive exhaustive search. As a result, we demonstrated how to take maximal advantage of more than half a century of research and development into rapid optimization. This makes it practical to run the whole analysis process as many times as desired on an entire high-throughput database. As a contrast, analyzing a high-throughput dataset of approximately 15 million samples using the Kalafut-Visscher step-finder[18] took around 40 hours and did not return useful results – in fact, manual editing labour was required so that reliable information could be extracted using this step-finder.

For these reasons, although these new analysis tools are quite simple, they significantly extend the limits of precision and applicability in the characterisation of discreteness and noise in molecular dynamics. The algorithms we describe here require minimal manual intervention. These tools therefore represent a major step towards the detailed, automated characterisation of the discrete behaviour of molecular machines. This step is required to fully exploit the promise of high-quality, high-throughput experiments on single and multiple-molecule machines. Furthermore, by speeding up the complete experiment-analysis cycle, we can facilitate the screening of multiple phenotypes or ranges of experimental conditions, so that novel structures and parameter values for mathematical models can be rapidly tested.

## Acknowledgements

MAL is funded through BBSRC/EPSRC grant number BBD0201901. RMB is funded through BBSRC grant number BBE00458X1. NSJ thanks the EPSRC and BBSRC. TB was funded by the EPSRC via the Life Sciences Interface Doctoral Training Centre. BCS is funded through an EMBO Long Term Fellowship and NIH grant R01GM066223. DMM is funded through NIH grant R01GM066223.

## Author Contributions

MAL implemented the analysis algorithms; MAL and BF performed simulations of molecular systems; MAL and NSJ designed the algorithms; YS carried out the flagellar motor experiments; BCS carried out the $F_1$-ATPase experiments; DMM modified $F_1$-ATPase for use in single molecule experiments; TB contributed preliminary data to test the developing techniques; MAL, RMB and NSJ wrote the paper.

## Competing Interests Statement

The authors declare no competing financial interests.

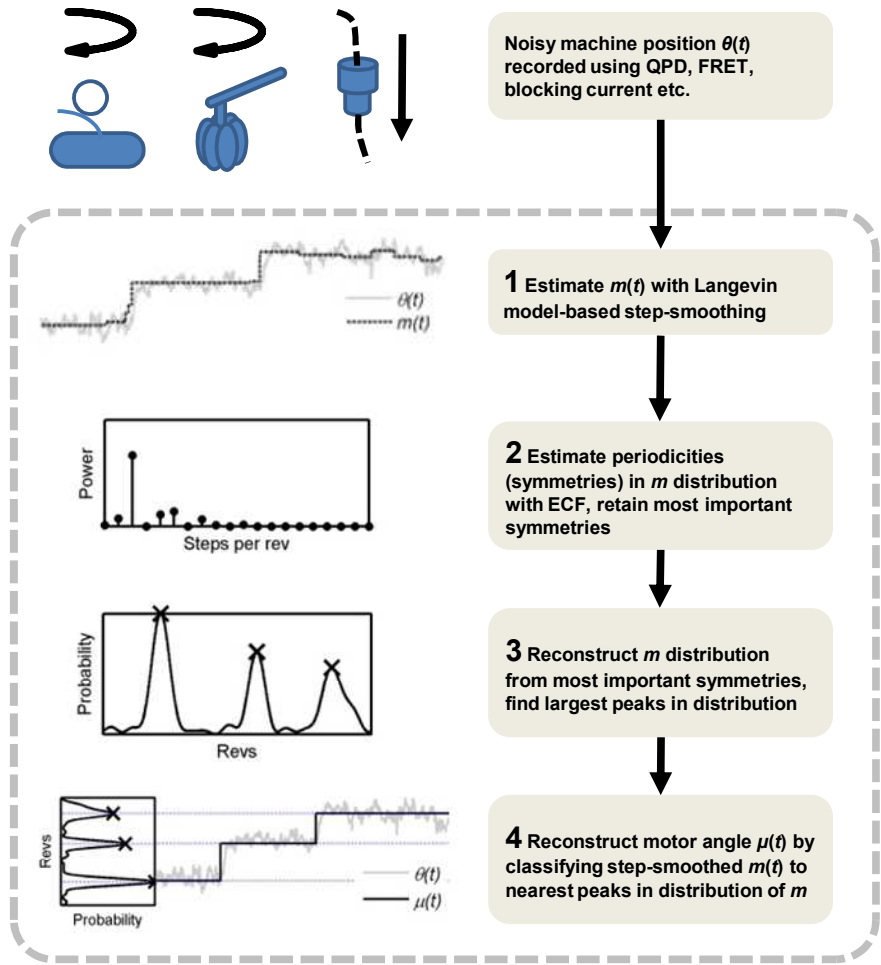

**Figures**

*Figure 1: Discrete state extraction process.* Experimental recording of noisy molecular machine position-time traces $\theta(t)$ are obtained, for example, bacterial flagellar motor angle captured by quadrant photodiode (QPD), charge-coupled device (CCD) video of flagellar-attached bead, $F_1$-ATPase angle using laser dark-field or Forster resonance energy transfer (FRET), or blocking current of DNA translocation through a hemolysin nanopore. Inside the dotted box is the analysis process applied to an example of $F_1$-ATPase. (1) Step-driven Langevin model fitted to series (algorithm L1-PWC-AR1, see text), and estimated time traces of machine positions $m(t)$ quickly and accurately recovered (see Figure 6). (2) Periodicities (symmetries) in distribution of $m$ estimated (using ECF-Bump algorithm, see text), and (3) most important symmetries retained and used to reconstruct distribution of $m$. (4) Classifying estimated $m(t)$ trace to nearest large distribution peaks (algorithm ML-Peaks, see text), an estimate of the true machine time-position trace $\mu(t)$ is recovered, from which secondary properties such as dwell times and dwell time distributions can be quantified.

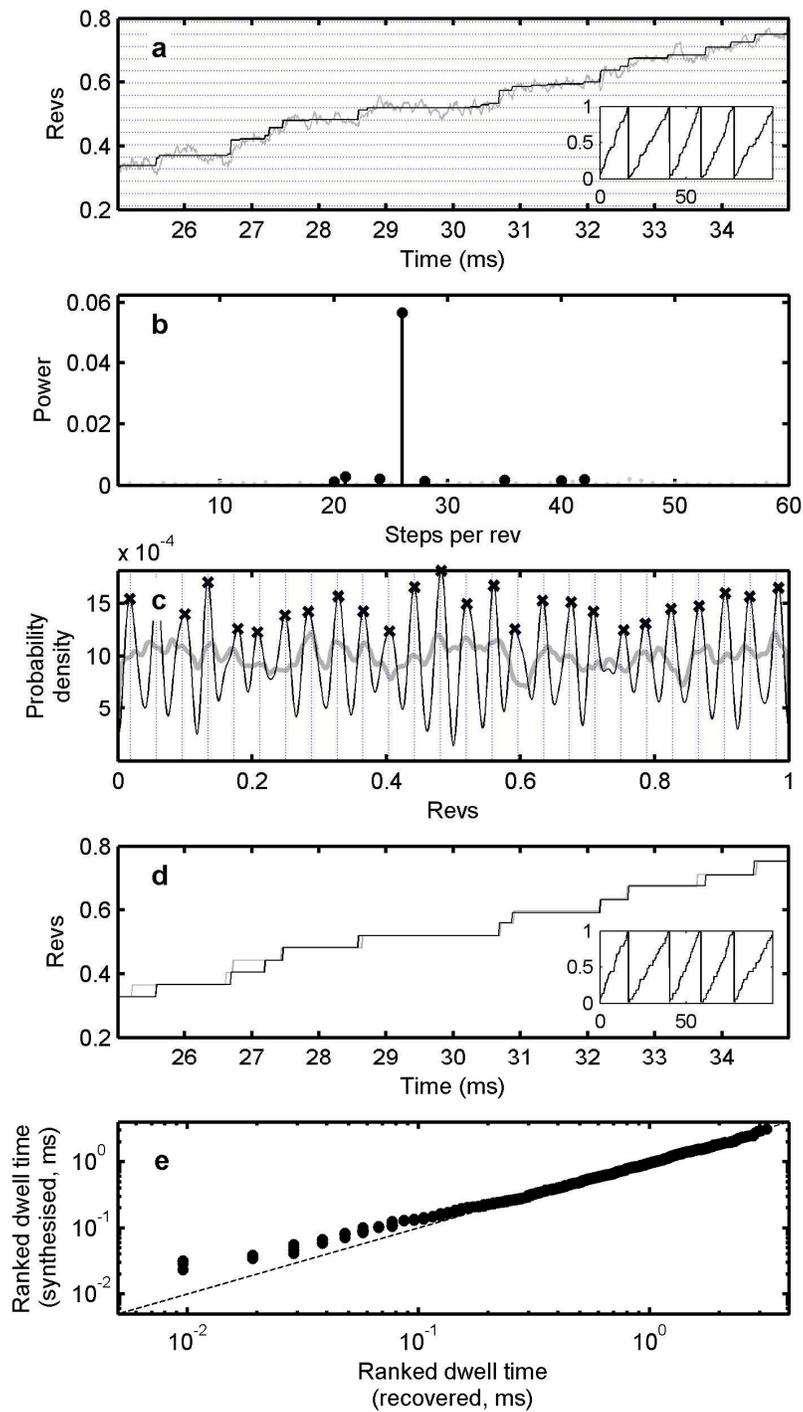

*Figure 2: Processing a synthetic time-angle trace.* State location extraction process for synthetic angle-time trace with 26 symmetric state locations. (a) A small segment of a noisy time trace showing rotation angles $\theta_t$ (gray), step-driven Langevin model estimates of motor positions $m_t$ (black), and state locations (blue dotted lines). Inset shows a longer segment of $m_t$. (b) Estimated symmetries (steps per revolution) in distribution of $m$ (gray), largest amplitude symmetries kept (black) and used to reconstruct distribution of $m$ (c). Reconstructed distribution of $m$ with all symmetries (gray), and with only the largest magnitude symmetries retained (black). (d) The estimated true motor angle time trace $\mu_t$ (black) is obtained by classifying the estimated $m_t$ trace to the largest distribution peaks, the black crosses in (c). (inset shows a longer segment of $\mu_t$). (e) Secondary properties such as dwell time distributions can be reliably quantified, here demonstrating a good fit to the true gamma dwell time distribution model (as the points lie close the dashed line). The horizontal axis shows the ranked logarithm of the estimated dwell times, and the vertical the ranked logarithm of synthesized dwell times drawn from a gamma distribution with the parameters used in the simulation.

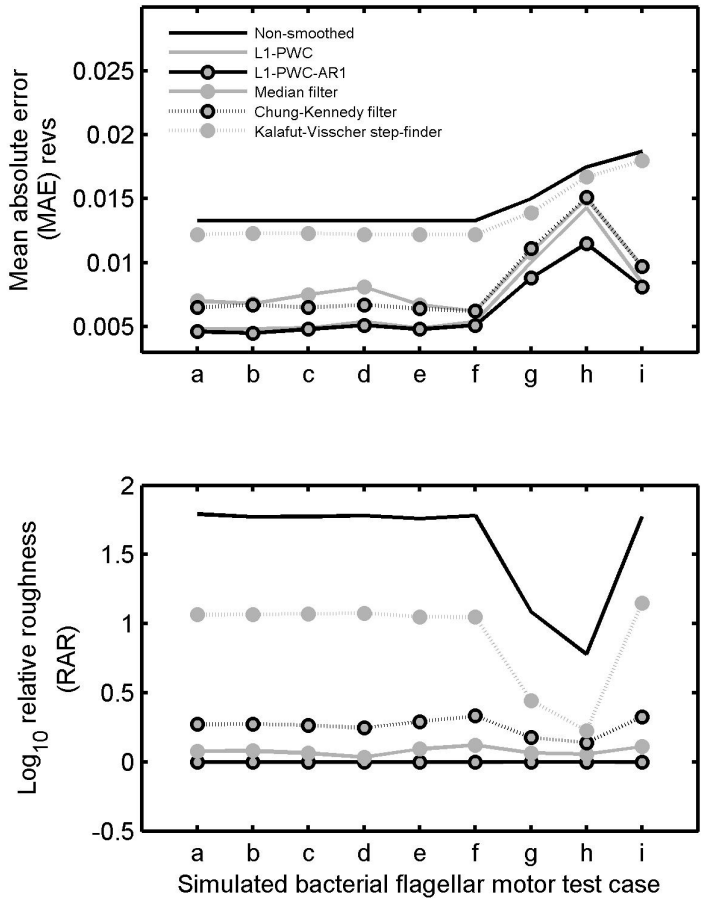

*Figure 3: Comparative performance of step-smoothing methods.* Top panel: mean absolute residual error, MAE (in revs, smaller is better) and bottom panel: (logarithm of) relative absolute roughness, RAR, closer to zero is better; for five step-smoothing methods and nine test cases of simulated bacterial flagellar motor rotation (see Supplementary Methods). RAR is the mean absolute difference between adjacent samples in the filter output $m_t$ divided by that in the true (simulated) motor position $\mu_t$. Values are averaged over five replications. Case (a) default test case has 26 equally-spaced state locations, exponentially-distributed dwell times, rotation 10 revs/sec, flagellar hook spring stiffness $\kappa = 100\ k_BT$/rad and coefficient of friction $\xi = 0.01\ k_BTs$. The other test cases differ from default case in one parameter. Case (b) has 20% dwell state asymmetry, (c) gamma distributed dwell times with shape $k = 2$, (d) gamma dwell times $k = 10$, (e) 30 dwell states, (f) 40 dwell states, (g) 50 revs/sec, (h) 100 revs/sec, (i) flagellar hook stiffness $\kappa = 50\ k_BT$/rad. Step smoothing algorithm parameters (see Supplementary Methods) are L1-PWC: $\gamma = 50$; L1-PWC-AR1: $\gamma = 1$, $P = 1$, $a_1 = 1 - \kappa \Delta t / \xi$; median filter window size is average dwell time in samples; Chung-Kennedy filter: filter length is half the dwell time in samples, analysis window $M = 16$ samples, weighting parameter $p = 0$. Note that for many test cases, the L1-PWC methods have indistinguishable RAR from the other step-smoothing methods, so that their curves lie almost on top of one another.

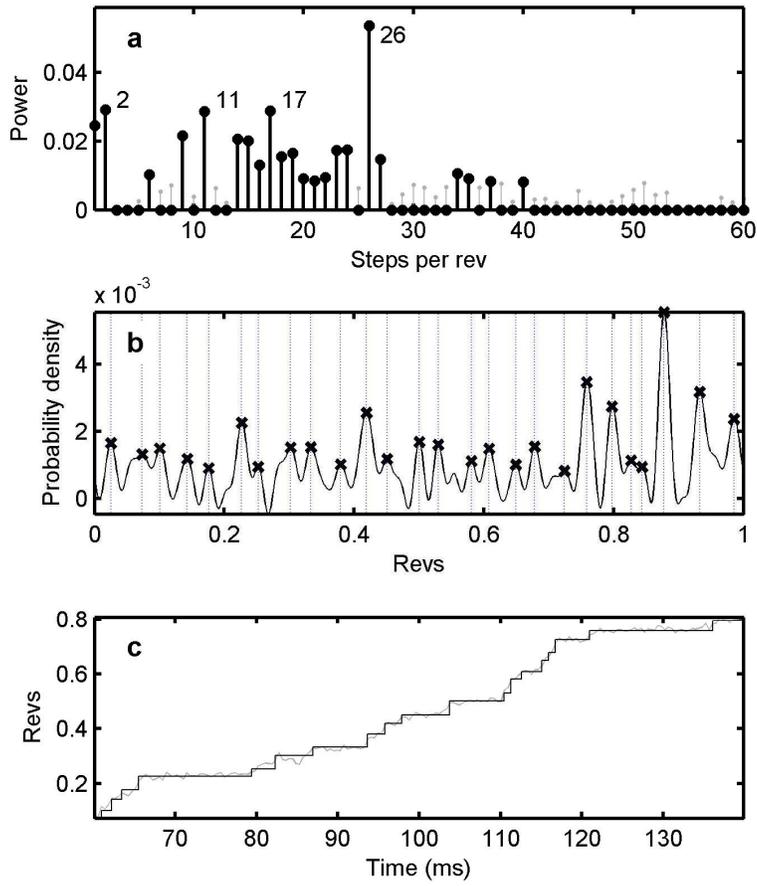

*Figure 4: Processing a single experimental flagellar motor time-angle trace.* An example of processing a single experimental *E. coli* flagellar motor time-angle trace (4.2s at 2.4kHz sampling rate) to extract discrete state locations. (a) Estimated symmetries in the distribution of the full, smoothed, 4.2s time series $m_t$ (gray), largest magnitude symmetries retained (black), showing the dominant 26-fold periodicity of this molecular machine. (b) The retained symmetries are used to reconstruct the distribution of $m$ (black line). The 26 largest peaks in this distribution (black crosses) represent the best estimate of the discrete state locations of the motor, and can be used to estimate the true time series of discrete state transitions. (c) A small segment of the recorded time trace showing rotation angles $\theta_t$ (gray) with classified estimates of motor positions $\mu_t$ following L1-PWC smoothing (black, see text for algorithm descriptions).

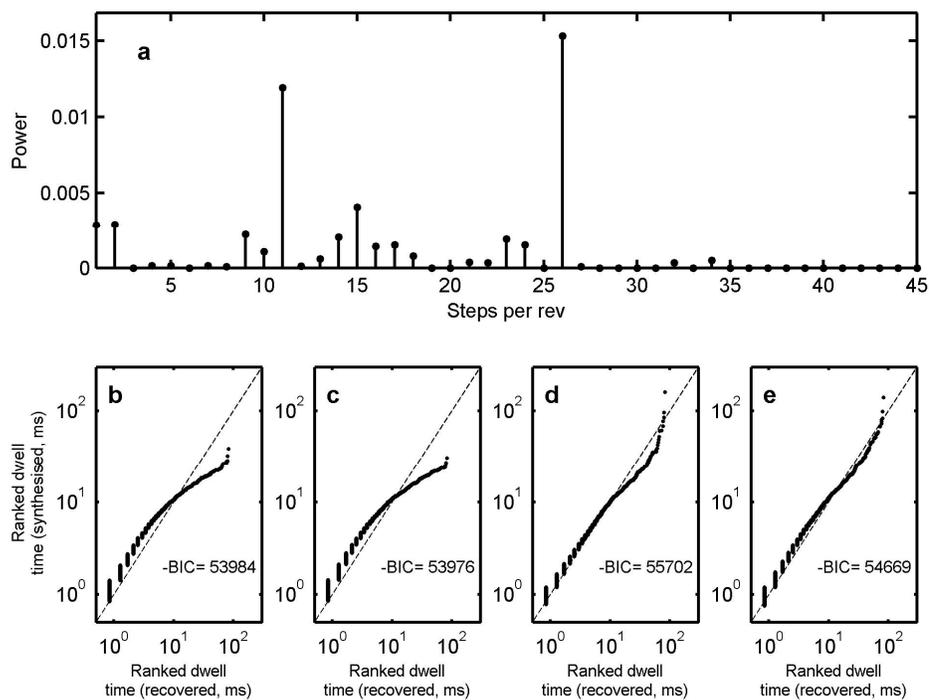

*Figure 5: Processing of all six traces recorded from the same motor of Figure 4.* Total time of these traces is 25.2s. (a) Periodicity (symmetry) analysis showing 26 states with superimposed 11-fold symmetry. The power spectrum is the median over all six power spectra obtained, including that shown in Figure 4(a). (b)-(e) Four different distribution models for dwell times. Each plot shows the ranked estimated dwell times (horizontal) versus the ranked synthesized dwell times drawn from the particular distribution model with the parameters fitted to the dwell times by a maximum likelihood procedure (vertical). Best fitting model is closest to diagonal (dashed line). Also shown are the (negative) Bayesian Information Criterion (BIC) values of the model fit (larger is better). (b) Exponential distribution (Poisson stepping), (c) gamma distribution, (d) Log-normal distribution, and (e) generalized Pareto distribution (here with parameters that define a heavy-tailed power-law distribution).

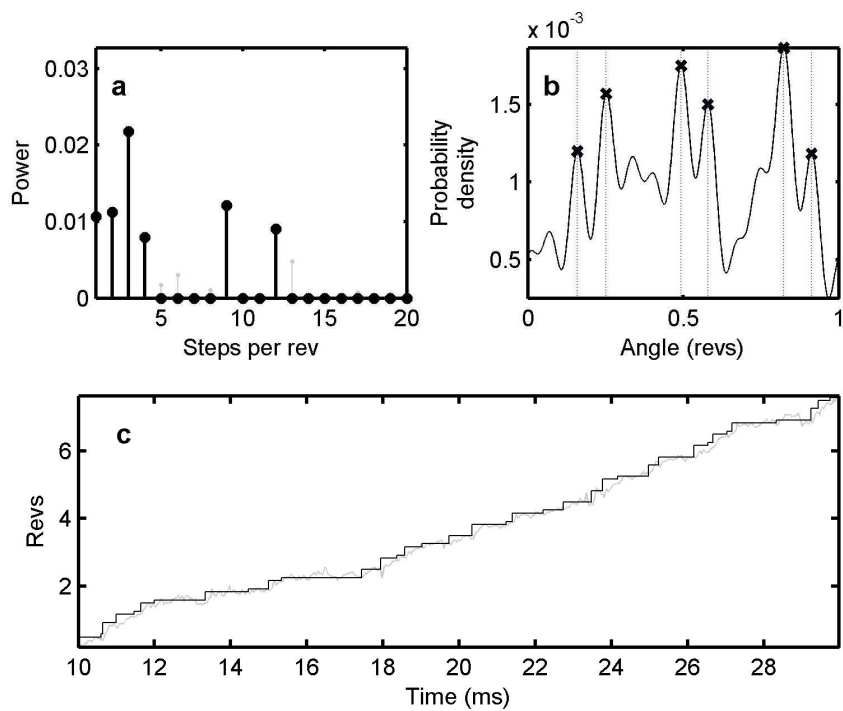

*Figure 6: Processing a single experimental $F_1$-ATPase time-angle trace.* Processing a 0.27s-long time-angle trace recorded at 30kHz to extract discrete state locations. (a) Estimated symmetries in the distribution of the full, smoothed, 0.27s time series $m_t$ (gray), largest amplitude symmetries retained (black), showing a combination of dominant three, nine and twelve-fold symmetries, which is consistent with six dwell locations. (b) Reconstructed distribution of states (black line) estimated using the six symmetries retained (between 1 and 20 steps per rev) after thresholding in (a). The six largest peaks in this distribution (black crosses) represent the best estimate of the discrete state locations of the motor, used to estimate the true time series of discrete state transitions (c, black). (c) A small segment of the trace showing rotation angles $\theta_t$ (gray) with classified estimates of motor positions $\mu_t$ following L1-PWC-AR1 smoothing (black, see text for algorithm descriptions).

## Methods

**Background: smoothing of correlated, step-like data.** *Step-smoothing* algorithms remove noise from step-like signals – that is, signals for which the underlying, noise-free time trace to be recovered consists of constant segments with step changes between segments. The literature on step-smoothing is large[7,8,11-14,18,23-26] but approaches can be roughly grouped according to how they process the data. *Sliding window* methods (such as the *running mean* or *running median filter*) consider a small set of contiguous samples and replace the central sample in that window with the output of an algorithm applied to the samples only in the window. The window is moved along the time series by one sample at a time so that each sample in the series is replaced. *Recursive* methods operate by successive subdivision or merging of constant regions. For example the *top-down* algorithm described in Kalafut *et al.*[18] starts by finding the location of the best single step for the whole signal. If a *stopping criterion* is not met, a best step location is found within each constant region either side of this step. This successive subdivision is repeated until the stopping criterion is reached. Conversely, *bottom-up* algorithms start with every sample as a putative constant region, and merge regions successively until a stopping criterion is reached. Finally, *global* methods consider all the samples simultaneously, either fitting a model to the time series or applying some kind of transformation. The range of algorithms is very diverse. For example, *Haar-wavelet de-noising*[24] decomposes the signal into a sum of step-like waveforms and the noisy details that fall below a given amplitude are removed from the sum. *Hidden Markov models* treat the underlying constant regions as unobserved states, obscured by noise, following each other in an unknown sequence, and attempt to find the state transition probabilities, observation noise probabilities, and most likely sequence of states from the signal[14].

All step-smoothing methods have advantages and disadvantages relative to each other, mostly consequences of the mathematical assumptions embodied in the algorithm. Most have algorithm parameters that must be set and the results will depend on the choice of those parameters. Without any guiding theoretical principles, there is always some unavoidable level of subjectivity. Sliding window methods cannot generally find constant regions larger than the window size; the window size therefore becomes a critical parameter. Recursive methods can find long constant regions, but require specification of the stopping criterion, and this may or may not be appropriate for the underlying trace which is of course unobservable in principle. For example, the top-down method described above produces different results on different length signals, an unintended side-effect the mathematical formulation of the particular stopping criterion. Global methods do not suffer from such problems but, for example, wavelet de-noising requires the selection of noise amplitude which is critical to the result. Hidden Markov models are an alternative, but they require a-priori choice of the number of states, and this number may actually be the subject of investigation.

The majority of step-smoothing algorithms make, usually for mathematical convenience, the (explicit or implicit) assumption that the observed noise is independent or specifically, *uncorrelated*. However, general molecular motion occurs in potential wells centred on each discrete state with correlated, random noise, and this type of motion is often modelled by *Langevin dynamics* which incorporates general time lag in the motion caused by elastic and friction forces[27] (for example, drag caused by a bead attached to the elastic flagellar hook[5]); there is an inevitable delay in the transitions. Step-smoothing under these circumstances is quite different from the step-smoothing problem as investigated in the existing statistical literature.

Additionally, many step-smoothing techniques are computationally onerous in practice, either because the number of intensive calculations scales poorly with the length of the signal, or because there is no way of knowing whether a solution constitutes the *globally optimal* set of steps other than by *exhaustive* (brute-force) computation. For example, most recursive methods require an exhaustive search for the globally optimal step location at each iteration[8,18], and, although there are some incidental computational efficiencies that can be exploited, for this reason such algorithms are intrinsically slow. For the hidden Markov model, there are no known algorithms that can evaluate the probabilities in their entirety without infeasible exhaustive computation so the best one can achieve is a solution that may be sub-optimal.

Given these considerations, we did not find any existing step-smoothing methods to be entirely satisfactory, and indeed our experiments on simulated data bear this out (see tables in the main text and Supplementary Methods).

**Background: finding modes (bumps) in distributions.** *Mode-finding* (also known as *bump-hunting*) is the activity of finding peak values of a distribution of a random variable. Typically the random variable represents time series of discrete states of a molecular complex or machine. If the states are distinct, then, in theory, the distribution will exhibit a clear set of "spikes" located at each unique state (see Figures 1, 2, 4 and 6). These separate states can then be identified from the distribution, and this requires an estimate of the distribution of the time series. Perhaps the simplest and most immediately accessible estimate is the histogram: divide the full range of the random variable into equal-sized bins and count the number of time series values that fall in each bin[15]. Assuming each bin is sufficiently small that there are at least two bins per discrete state, the maxima of the histogram can locate estimates of the discrete states of the molecular machine.

However, there are problems with histograms because the separation between states is usually not known in advance, molecular motion is generally obscured by thermal and observational noise, and we only have a finite number of samples. As the precision of the peak location estimates is increased by decreasing the bin size, the bins become more sensitive to these confounding factors and spurious peaks start to emerge. Similarly, increasing the bin size to make the peaks less spurious decreases the precision of the peak location estimates and may cause two or more states to become inseparable. Also, location estimates will tend to be sensitive to the choice of bin edges, and non-equal bin sizes appear mainly to introduce complications with no special advantages[15].

Averaging over different bin edges has been proposed (*average shifted histograms*), but it can be shown that this is a special case of *kernel density estimation*, where a smooth function (usually with one mode – *unimodal*) of fixed width is centred on each time series sample, and the distribution at every location is estimated as the equally weighted sum of all these functions[15]. Kernel density estimates are an improvement over histograms because smooth distribution estimates stabilize peak-finding in the presence of noise. The choice of kernel width is however crucial, because, if the width is too small, spurious peaks will emerge, and if too large, distinct states may become merged erroneously.

Incorporating additional information about the discrete states may lead to improvements. If the number of states is known in advance, *mixture modelling* may be used to find the best combination of a weighted sum of component distributions with arbitrary locations and widths[16]. With unimodal component distributions, one distribution will be, ideally, located at each individual state. The main difficulty with mixture modelling is that the simultaneous estimation of the component locations and widths is not a *convex problem*, so that we cannot guarantee that any solution we find is the best one, and the computations quickly become onerous as the number of components grows. Similar issues apply to *k-means clustering* which attempts to cluster the time series samples into a given number of states by minimizing the total distance between all samples assigned to each cluster. An algorithm for finding a solution exists[16], and although the problem is simpler than mixture modelling because only the state locations need to be estimated, the results can often depend quite sensitively on the initial choice of assignments required to start the search for a solution.

In estimating discrete states therefore, as with step-smoothing, existing bump-hunting approaches described above are problematic (see Supplementary Tables 1 and 2).

**Physically-based extraction of quasi-periodic discrete state dynamics.** The above discussions highlight some of the limitations of current step-smoothing and bump-hunting approaches. Our main observation is that these techniques have been developed to solve different problems than the one with which we are actually concerned. In detecting molecular dynamics we have a lot of additional information that can serve as constraints on the algorithms to produce more efficient results. In particular, the time series can be highly correlated. Thus, general bump-hunting, which throws away time in order to estimate the distribution of states, will not be as effective as a more specific method that takes time into account. The steps may also have some known geometric relationship to each other. For instance, the steps may have several dominant symmetries that are naturally represented in a Fourier basis.

Our point of departure is the classical *autoregressive* (AR) smoothing filter, that assumes the noisy angle-time trace $\theta_t$ is generated by a recursive equation $\theta_{t+1} = a\theta_t + \varepsilon_t$, where $a$ is a feedback constant and $\varepsilon_t$ a white noise process. A plausible model for the typical experimental setup involving damping and potential energy storage is Brownian motion in potential well $d\theta = \rho(\mu - \theta)dt + \sigma dW$, where $\mu$ is the machine position, $W$ a Wiener process, $\sigma$ the diffusion coefficient, and $\rho$ the drift coefficient. This model can be simulated with the discrete-time equivalent $\theta_{t+1} = a\theta_t + (1-a)\mu_t + \varepsilon_t$, for the constant $a$ which can be found from the autocorrelation at one sample lag of the time-angle trace (see Supplementary Methods). Then the goal of the filter is to find the best approximation of the machine position trace $m_t$ observing only $\theta_t$. This problem can be solved by finding the $m_t$ for the full time range $t$ that minimizes, for example, the sum of square errors $\Sigma_t \varepsilon_t^2$. In general, this global optimization problem cannot be solved without imposing some additional conditions on $m_t$. Therefore, *step-smoothness constraints* have been introduced, requiring that $m_t$ is constant in time except at the points where the motor changes state, i.e. at the step time instants[23].

On the surface, this appears to be an intractable optimization problem because there are an unknown number of step time instants that can occur anywhere in the full range of times $t$, so that finding these instants requires brute-force testing of every possible combination of time instants for the presence of steps. However, use can be made of recent theoretical innovations. These apply to the particular step-smoothness constraint penalizing the sum of absolute differences between successive instants of $m_t$. The theoretical result shows that if there exist only a finite number of steps, solving the resulting optimization problem finds the correct positions of all the steps[28]. The resulting optimization problem is convex (in this case, the sum of the model fit error with the step constraint has only one minimum with respect to variation in the unknown $m_t$):

$$\hat{m}_t = \arg\min_{m_t} \sum_{t=P+1}^{T}\left(\theta_t - \sum_{i=1}^{P} a_i \theta_{t-i} - m_t\right)^2 + \gamma \sum_{t=2}^{T}|m_t - m_{t-1}| \quad (1)$$

(We call this algorithm L1-PWC-ARP, and with all $a_i$ zero, algorithm L1-PWC, see Supplementary Methods and Kim *et al.* (2009) for derivations of this approach). Here, $T$ is the length of the time trace, and the $a_i$ are determined from a biophysical model, or from analysis of the time trace (see Supplementary Methods). As the constraint constant $\gamma$ increases, larger weight is placed on the step-smoothing, so that the resulting $m_t$ is increasingly smooth, at the expense of increasing the sum of squares error. This *quadratic programming* problem has the desirable property that a guaranteed globally optimal solution can be obtained using standard optimization techniques with computing time and resources that increase very slowly with increasing data size[29]. There are several methods for selecting the optimal $\gamma$, including plotting the first summation term of (1) against the second summation term over a wide range of values and choosing the value at which both terms change the least. In this paper, for the simulations we choose the value that produces the optimal result on simulated data (see Supplementary Methods). For real data, we select the default $\gamma = 1$ that places equal weight on approximation error (first term) and smoothness (second term).

Given the smoothed time trace, we need to identify the most likely locations of the discrete molecular states and their periodicities (*symmetries*) and this requires an estimate of the distribution of states. Our answer to the problems of peak-finding with histograms or kernel density approaches is to estimate the distribution *directly* in the Fourier domain (*empirical characteristic function*, ECF) estimated by calculating:

$$P(f_j) = \frac{1}{T} \sum_{t=1}^{T} \exp(if_j m_t) \quad (2)$$

for the $K$ symmetries of interest $f_1, f_2 \ldots f_K$ (see Supplementary Methods for a derivation of this expression). In this domain, periodicity is naturally represented by only a few dominant non-zero coefficients in the power spectrum (the ECF is a *sparse representation* for periodic distributions[30]) and we avoid problematic histogram (or other[15]) distribution estimates altogether. Since we only have a finite amount of data and there are experimental confounds, the power spectrum will be the sum of sampling effects and experimental noise and the actual symmetries of the molecular machine. Then the problem is to detect which symmetries are genuine, and which are due to sampling noise and other artefacts.

In this case, the statistical theory of *nonlinear threshold estimation* provides us with the guarantee that (assuming physically reasonable smoothness-like constraints on the distribution), this detection problem is solved by simply setting the coefficients for all symmetries with power below a certain threshold to zero, or, which is equivalent, retaining only a fraction $\varphi$ of the largest power symmetries[30]. Setting coefficients above an upper frequency limit to zero is similar to kernel density estimation of the distribution that smoothes away finite sampling effects[15] (see Supplementary Methods, algorithm ECF-Bump). Thus, we can accomplish both symmetry detection and finite sample effect smoothing by Fourier coefficient selection. Here, we choose the smoothing frequency parameter to be sufficiently large that we can always capture symmetries of interest. The nonlinear threshold fraction $\varphi$ is chosen such that under simulated conditions, we can always recover all the known symmetries.

The noise-free distribution of states $p(m)$ is then approximated by applying the inverse Fourier transform to the ECF coefficients:

$$p(m) \approx \sum_{j=-K}^{K} \exp(-if_j m) \overline{P}(f_j) \quad (3)$$

Having obtained the distribution, the peaks represent the discrete states, from which, for example, the step-like time traces of states $\mu_t$ can be recovered. Knowing the dominant symmetry $N$, of the distribution, we can expect $N$ peaks. At each time step $t$ we wish to determine the state of the machine $\mu_t$. This is a *statistical classification problem*: we wish to find the optimum peak to assign to each time step. A statistically consistent solution to this problem is the *maximum likelihood* approach obtained by assigning the step-smoothed time series $m_t$ to the nearest peak. This is the solution to the problem:

$$\hat{\mu}_t = \arg\min_{\mu_n: n=1,2\ldots N} |m_t - \mu_n| \quad (4)$$

where $\mu_n$ are the locations of the $N$ largest peaks in $p(m)$. Finally, the time traces of states can be used to estimate the time spent in each state (the *dwell times*), and models for the distribution of these dwell times can be found and compared.

**Experimental methods.** Bacterial flagellar motor time-angle traces were obtained by video microscopy of 200nm fluorescent beads attached to the truncated flagellar filaments of surface-immobilized *E. coli* chimaeras. Please see Sowa *et al.* (2005) for further details. ATPase time-angle traces were obtained by dark field microscopy of a 60nm gold bead attached to the gamma subunit of His$_6$-tagged *S. Cerevisiae* F$_1$-ATPase via a streptavidin-biotin linker. The molecule was immobilized onto a Ni$^{2+}$NTA surface, at an ATP concentration of 30μM. Images were captured with a Photron PCI1024 at a frame and shutter rate of 30kHz, and the bead position was calculated using the gaussian mask algorithm described in Thompson *et al.*[31]

**References**

1. Nelson, P.C., Radosavljevic, M. & Bromberg, S. *Biological physics: energy, information, life*, xxvi, 598 p. (W.H. Freeman and Co., New York, 2004).
2. Raj, A. & van Oudenaarden, A. Nature, Nurture, or Chance: Stochastic Gene Expression and Its Consequences. *Cell* **135**, 216-226 (2008).
3. Mori, T., Vale, R.D. & Tomishige, M. How kinesin waits between steps. *Nature* **450**, 750-U15 (2007).
4. Kitamura, K., Tokunaga, M., Iwane, A.H. & Yanagida, T. A single myosin head moves along an actin filament with regular steps of 5.3 nanometres. *Nature* **397**, 129-134 (1999).
5. Sowa, Y. et al. Direct observation of steps in rotation of the bacterial flagellar motor. *Nature* **437**, 916-919 (2005).
6. Selvin, P.R. & Ha, T. *Single-molecule techniques: a laboratory manual*, vii, 507 p. (Cold Spring Harbor Laboratory Press, Cold Spring Harbor, N.Y., 2008).
7. Carter, B.C., Vershinin, M. & Gross, S.P. A comparison of step-detection methods: How well can you do? *Biophysical Journal* **94**, 306-319 (2008).
8. Kerssemakers, J.W.J. et al. Assembly dynamics of microtubules at molecular resolution. *Nature* **442**, 709-712 (2006).
9. Arce, G.R. *Nonlinear signal processing: a statistical approach*, xx, 459 p. (Wiley-Interscience, Hoboken, N.J., 2005).
10. Meacci, G. & Tu, Y. Dynamics of the bacterial flagellar motor with multiple stators. *Proceedings of the National Academy of Sciences of the United States of America* **106**, 3746-3751 (2009).
11. Chung, S.H. & Kennedy, R.A. Forward-Backward Nonlinear Filtering Technique for Extracting Small Biological Signals from Noise. *Journal of Neuroscience Methods* **40**, 71-86 (1991).
12. Fried, R. On the robust detection of edges in time series filtering. *Computational Statistics & Data Analysis* **52**, 1063-1074 (2007).
13. Pawlak, M., Rafajlowicz, E. & Steland, A. On detecting jumps in time series: Nonparametric setting. *Journal of Nonparametric Statistics* **16**, 329-347 (2004).
14. Jong-Kae, F. & Djuric, P.M. Automatic segmentation of piecewise constant signal by hidden Markov models. in *Statistical Signal and Array Processing, 1996. Proceedings., 8th IEEE Signal Processing Workshop on (Cat. No.96TB10004* 283-286 (1996).
15. Silverman, B.W. *Density estimation for statistics and data analysis*, ix, 175 p. (Chapman & Hall/CRC, Boca Raton, 1998).
16. Hastie, T., Tibshirani, R. & Friedman, J.H. *The elements of statistical learning : data mining, inference, and prediction : with 200 full-color illustrations*, xvi, 533 p. (Springer, New York, 2001).
17. Clarke, J. et al. Continuous base identification for single-molecule nanopore DNA sequencing. *Nature Nanotechnology* **4**, 265-270 (2009).
18. Kalafut, B. & Visscher, K. An objective, model-independent method for detection of non-uniform steps in noisy signals. *Computer Physics Communications* **179**, 716-723 (2008).
19. Block, S.M., Blair, D.F. & Berg, H.C. Compliance of Bacterial Flagella Measured with Optical Tweezers. *Nature* **338**, 514-518 (1989).
20. Junge, W., Sielaff, H. & Engelbrecht, S. Torque generation and elastic power transmission in the rotary F0F1-ATPase. *Nature* **459**, 364-370 (2009).
21. Thomas, D.R., Francis, N.R., Xu, C. & DeRosier, D.J. The three-dimensional structure of the flagellar rotor from a clockwise-locked mutant of Salmonella enterica serovar typhimurium. *Journal of Bacteriology* **188**, 7039-7048 (2006).
22. Leake, M.C. et al. Stoichiometry and turnover in single, functioning membrane protein complexes. *Nature* **443**, 355-358 (2006).
23. Kim, S.J., Koh, K., Boyd, S. & Gorinevsky, D. L1 Trend Filtering. *SIAM Review* **51**, 339-360 (2009).
24. Cattani, C. Haar wavelet-based technique for sharp jumps classification. *Mathematical and Computer Modelling* **39**, 255-278 (2004).
25. Hou, Z.J. & Koh, T.S. Robust edge detection. *Pattern Recognition* **36**, 2083-2091 (2003).
26. Smith, D.A. A quantitative method for the detection of edges in noisy time-series. *Philosophical Transactions of the Royal Society of London Series B-Biological Sciences* **353**, 1969-1981 (1998).
27. Becker, O.M. *Computational biochemistry and biophysics*, xii, 512 p. (M. Dekker, New York, 2001).
28. Strong, D. & Chan, T. Edge-preserving and scale-dependent properties of total variation regularization. *Inverse Problems* **19**, S165-S187 (2003).
29. Boyd, S.P. & Vandenberghe, L. *Convex optimization*, xiii, 716 p. (Cambridge University Press, Cambridge, UK ; New York, 2004).
30. Candes, E.J. Modern statistical estimation via oracle inequalities. *Acta Numerica* **15**, 257-326 (2006).
31. Thompson, R.E., Larson, D.R. & Webb, W.W. Precise nanometer localization analysis for individual fluorescent probes. *Biophysical Journal* **82**, 2775-2783 (2002).

## Supplementary Methods

**Physically-based step-smoothing, quasi-periodic bump-hunting and distribution estimation.** We seek algorithms for step-smoothing, bump-hunting and distribution estimation that incorporate physical knowledge of molecular conformational dynamics. In the following, we have a time-position trace $\theta_t$, $t = 1, 2 \ldots T$ obtained from experimental measurements of molecular dynamics. The series $\mu_t$ is the unknown series of positions corresponding to conformations of the molecular machine to be determined, and the series $m_t$ is an approximation to $\mu_t$. We require that any algorithm results can be obtained with reasonable computational cost, and that they are guaranteed to converge on the globally optimal solution.

*Algorithm L1-PWC*: *L1-regularized, global step-smoothing with independent noise*. The smoothed estimate is constructed by minimizing the following *negative log posterior* (NLP) cost function with respect to $m_i$:

$$NLP = \sum_{t=1}^{T}(\theta_t - m_t)^2 + \gamma \sum_{t=2}^{T}|m_t - m_{t-1}| \tag{1}$$

The implicit physical model is in the form $\theta_t = \mu_t + \varepsilon_t$, where $\mu_t$ is a time series consisting of constant segments with abrupt jumps (steps), and $\varepsilon_t$ is a time series of independent Gaussian noise. The problem is to find the series $m_t$ that approximates the true time-angle step trace $\mu_t$ which consists of the piecewise constant steps buried in the noise, but that is simultaneously a good approximation to the recorded time series $\theta_t$. The first term in the NLP represents the error (negative log-likelihood) of the smoothed approximation $m_t$. The second term represents the total absolute difference between consecutive approximation samples (in the Bayesian interpretation this is the negative log prior). When the penalty (*regularization*) term $\gamma = 0$, the solution becomes $m_t = \theta_t$ and no smoothing occurs. Thus, the useful behaviour of this algorithm occurs when $\gamma$ increases, so that increasing weight is placed on minimizing the second term at the expense of the first. At the extreme when $\gamma \to \infty$, the solution becomes $m_t = \frac{1}{T}\sum_{t=1}^{T}\theta_t$, i.e. all approximation samples take on the mean of the recorded time series. Assuming that there are only a small number of steps in $\mu_t$ amounts to imposing a *sparsity condition* on $\sum_t |m_t - m_{t-1}|$, that is, only a few terms in this expression are non-zero[1]. Under this condition it is possible to recover an approximation to $\mu_t$ that finds the true locations of the steps[2]. Thus, increasing $\gamma$ forces most of the differences between consecutive samples of $m_t$ to zero. This NLP cost function is *convex* and *quadratic* so that the optimal approximation can be obtained by minimizing NLP with respect to $m_t$ using standard quadratic programming techniques[3]. This algorithm is similar to optimal piecewise linear smoothing[4], and is in other circles known as discrete *total variation denoising*[2]. In the main text, we demonstrate the use of this algorithm to approximate step-like motion in experimental bacterial flagellar motor time-angle traces where the sampling rate is sufficiently low that the stepping is effectively instantaneous.

*Algorithm L1-PWC-ARP*: *L1-regularized global step-smoothing with known correlated noise*. This algorithm is an adaptation of the L1-PWC algorithm that incorporates a general, discrete-time Langevin model within the filter structure. It minimizes the following cost function:

$$NLP = \sum_{t=P+1}^{T}\left(\theta_t - \sum_{i=1}^{P}a_i\theta_{t-i} - m_t\right)^2 + \gamma \sum_{t=2}^{T}|m_t - m_{t-1}| \tag{2}$$

Here the implicit model is $\theta_t = \sum_{i=1}^{P}a_i\theta_{t-i} + \mu_t + \varepsilon_t$ which captures general Gaussian, linear, discrete-time stochastic dynamics. The real-valued coefficients $a_i$ represent the discrete-time feedback of past values of the recorded time series on the current value. This model incorporates the special case of discrete-time Langevin dynamics with linear drift and diffusion terms, in widespread use as models for general molecular dynamics[5]. The forcing term $\mu_t$ consists of piecewise constant segments with jumps. Again, minimizing this cost function is convex and quadratic and can be solved for $m_t$ using a standard quadratic programming algorithm. The underlying constant step approximation is then recovered as $m_t / \left(1 - \sum_{i=1}^{P}a_i\right)$. In particular, in the main text we demonstrate use of the special case with $P = 1$ (L1-PWC-AR1) to capture Langevin motion in an experimental F$_1$-ATPase angle-time trace:

$$NLP = \sum_{t=2}^{T}(\theta_t - a_1\theta_{t-1} - m_t)^2 + \gamma \sum_{t=2}^{T}|m_t - m_{t-1}| \tag{3}$$

*Algorithm ECF-Bump*: *Nonparametric bump-hunting*. The *characteristic function* is an alternative representation of the distribution $p(m)$ of the step-smoothed time series of molecular states:

$$P(f) = \int_{-\infty}^{\infty} \exp(ifm)p(m)\,dm \tag{4}$$

In this context, $p(m)$ is unknown. However, $P(f)$ can be estimated from the time series $m_t$, using the *empirical characteristic function* (ECF):

$$P(f_j) \approx \frac{1}{T} \sum_{t=1}^{T} \exp(if_j m_t) \qquad (5)$$

Here, the ECF is evaluated over a set of chosen frequencies $f_j$, $j = 1, 2 \ldots K$. The distribution function $p(m)$ can be reconstructed from the coefficients $P(f_j)$, and covers the range $[0, 2\pi]$:

$$p(m) \approx \sum_{j=-K}^{K} \exp(-if_j m) \overline{P(f_j)} \qquad (6)$$

where the overbar denotes complex conjugation (and $f_{-j} = -f_j$ to ensure that $p(m)$ is a real probability). Each frequency $f_j$ corresponds to a potential symmetry (periodicity) of the molecular machine. In the special case of a rotary machine it corresponds to the number of steps per complete revolution. This characteristic function is closely related to the Fourier transform of the distribution of the time series. Thus, the $P(f_j)$ can be interpreted as *Fourier coefficients* of the distribution function, and are calculated over a range of experimentally relevant symmetries $f_j$.

The advantage of this representation of the distribution is that it is usually the case for a molecular machine that it has repeating molecular structures and so undergoes motion in a series of repeating steps. This implies that the distribution of states is both bounded and periodic, so the reconstruction converges on the exact distribution extremely rapidly as more frequency components are introduced ($K$ increases). This is because the Fourier transform is a *sparse representation*[6] for smooth, bounded, periodic functions. Sparse representations have the desirable property that only a few of the coefficients are large, and these are the ones that contribute most to the shape of the distribution. The rest of the coefficients will fluctuate due to experimental noise and contribute little, if anything. Thus, considering the coefficients as the sum of true molecular machine symmetries and experimental artefacts and distortions, we can apply *nonlinear thresholding* by ranking coefficients by their power $|P(f_j)|^2$ and retaining a small fraction $\varphi$ of the largest coefficients. It has been more recently shown that this simple procedure for recovering the noisy distribution is statistically optimal if the distribution is bounded, smooth and periodic[6], for an appropriate choice of $\varphi$ related to the amount of experimental noise. This nonlinear thresholding procedure contrasts with *linear thresholding* where only a certain number of the *lowest frequency* components are retained (in this context, kernel density smoothing is effectively a linear thresholding operation, and it therefore gives us no opportunity to retain high frequency components if they are actually important). Because we only have a small number of samples of $m_t$, the higher frequency coefficients fluctuate due to statistical finite sample effects. Therefore, we can also apply linear thresholding by retaining only those coefficients below a threshold. In practice however, nonlinear thresholding tends to remove most irrelevant high frequency components anyway.

Analysis of the characteristic frequency domain reveals important information about the symmetries of the molecular machine, since probability calculations with combinations of random variables become very simple in this domain. If we change the scale of a random variable $X$ by multiplying it by a scaling factor $\sigma$ and adding a constant $\mu$, the new random variable $Y = \sigma X + \mu$ has the following characteristic function:

$$P_Y(f) = E[\exp(if\{\sigma X + \mu\})] = \exp(if\mu) P(\sigma f) \qquad (7)$$

Hence, shifting the location of the peak of a distribution corresponds simply to multiplying the characteristic function by $\exp(if\mu)$. Similarly, increasing the spread of the peak corresponds to decreasing the width of the characteristic function. In the case when the distribution is composed of superposed bumps of width scaled by $\sigma$ and with period $N$, we have that:

$$P(f) = P_{\text{bump}}(\sigma f) \frac{1}{N} \sum_{n=0}^{N-1} \exp(if\mu_n) = P_{\text{bump}}(\sigma f) \frac{1}{N} \sum_{n=0}^{N-1} \exp\left(\frac{i2\pi nf}{N}\right) = P_{\text{bump}}(\sigma f) \frac{1}{N}\left(\frac{1 - \exp(i2\pi f)}{1 - \exp(i2\pi f/N)}\right) \qquad (8)$$

For the purposes of analysis, the *power* of the characteristic function is usually more convenient to work with than the characteristic function, and we are interested in whole integer symmetries $f$ only:

$$|P(f)|^2 = |P_{\text{bump}}(\sigma f)|^2 \frac{1}{N^2}\left(\frac{1 - \cos(2\pi f)}{1 - \cos(2\pi f/N)}\right) = |P_{\text{bump}}(\sigma f)|^2 \times \begin{cases} 1 & f = kN \\ 0 & \text{otherwise} \end{cases} \qquad (9)$$

where $k = 1, 2, \ldots$ is the multiple and $\sigma$ is the spread due to noise of the bump at each molecular state. This shows that the power of the characteristic function for a periodic distribution consists of a series of non-zero coefficients at integer multiples of the period $N$, the rest are zero. The absolute square magnitude of these non-zero coefficients is proportional to the absolute square magnitude of the characteristic function of the bump distributions, so that the spikes are attenuated in magnitude as the multiple increases. The characteristic function of many well-known distributions can be found exactly. For example, if the bump distributions are Laplacian, since $P_{\text{bump}}(f) = 1/(1 + \sigma^2 f^2)$, we obtain, at the peaks:

$$|P(f = kN)|^2 = \left(1 + [\sigma k N]^2\right)^{-2} \qquad (10)$$

Similarly, for Gaussian bumps $|P(f = kN)|^2 = \exp(-\sigma^2 [kN]^2)$. Qualitatively, as the noise spread increases, the non-zero coefficients in the characteristic function diminish in magnitude. Therefore, the sharper the bumps in the distribution, the easier it will be to identify the period above the background of finite sample variability and experimental artefacts.

In some cases, there will be an arrangement of molecular states that has more than one superimposed period, in general, a set of $Q$ different periods $N_q$ for $q = 1, 2 \ldots Q$ which are not multiples of each other. In this case, the characteristic function power will be:

$$|P(f=0)|^2 = 1, \quad |P(f=kN_q)|^2 = |P_{\text{bump}}(\sigma k N_q)|^2 N_q^2 \left(\sum_{r=1}^{Q} N_r\right)^{-2} \tag{11}$$

and $|P(f)|^2 = 0$ otherwise. Thus the power of the characteristic function has a series of non-zero coefficients at every integer multiple of each of the $Q$ constituent periods. The non-zero coefficients for period $N_q$ will have absolute square magnitude proportional to $N_q^2 \left(\sum_{r=1}^{Q} N_r\right)^{-2}$, so that larger periods have larger absolute square magnitude. Again, the non-zero coefficients will be attenuated in magnitude by the characteristic function of the bump distribution, and this will typically decrease faster with increasing noise spread. Therefore, superimposed symmetries in the molecular machine can be readily detected from analysis of the largest peaks in the power spectrum.

*Algorithm ML-Peaks*: *maximum likelihood reconstruction of discrete state time trace from distribution peaks*. Using algorithm ECF-Bump and applying the inverse Fourier transform to the coefficients $P(f_j)$, we can reconstruct the distribution $p(m)$ of molecular states. This distribution may have some small peaks that are due to finite sampling effects or inaccuracies in the reconstruction of the molecular state time trace $m$. However, the largest peaks are associated with the most dominant, and also most likely, positions of the molecular states. Thus, if the known dominant symmetry is $M$ steps per revolution, this information can be used to select the $M$ largest peaks in the distribution as the dominant discrete molecular state dwell locations. Having located these peaks, the step-smoothed time trace $m_t$ can be used to reconstruct the true step-like conformational state signal $\mu_t$ by classification of each of the $m_t$ to the nearest retained peak in the distribution. This classification is the *maximum likelihood* reconstruction of $\mu_t$ (see main text).

**Simulations of molecular machines**. Here we describe a model of Brownian motion in a potential well for periodic stepping motion of a molecular machine with frictional drag and elastic energy storage. We set up a simple linear stochastic differential equation (SDE) for a typical experiment. We measure the machine conformation through a small load attached to the machine whose observed position is $\theta$. The spring potential of the structure attaching the machine to the load is:

$$U(\theta) = \frac{1}{2}\kappa\theta^2 \tag{12}$$

where $\kappa$ is the spring stiffness constant. The load causes drag on the machine, represented using the linear friction model:

$$F(\dot{\theta}) = \xi\dot{\theta} \tag{13}$$

where $\xi$ is the friction coefficient. Assuming that the machine executes random motion about the equilibrium position $\theta = 0$, a Langevin equation of motion for the experiment can be written as:

$$M\ddot{\theta} + F(\dot{\theta}) + \nabla U(\theta) = \sqrt{2k_BT\xi}\,\varepsilon \tag{14}$$

where $k_B$ is Boltzmann's constant, $T$ is temperature, $M$ is the machine mass, and $\varepsilon$ is an independent, Gaussian random driving force with mean zero and unit standard deviation. Because the ratio $M/\xi$ is very small, the inertial term $\ddot{\theta}$ is negligible and we obtain the equations of motion:

$$F(\dot{\theta}) = -\nabla U(\theta) + \sqrt{2k_BT\xi}\,\varepsilon$$
$$\xi\dot{\theta} = -\kappa\theta + \sqrt{2k_BT\xi}\,\varepsilon \tag{15}$$

Including the average machine position $\mu(t)$ we obtain the following stochastic differential equation:

$$d\theta = \frac{\kappa}{\xi}(\mu - \theta)dt + \sqrt{\frac{2k_BT}{\xi}}dW \tag{16}$$

This is an *Ornstein-Uhlenbeck* process with mean $\mu$, drift $\rho = \kappa/\xi$, diffusion $\sigma = \sqrt{2k_BT/\xi}$ and Wiener process $W(t)$. Focusing on the motion of one step to the position $\mu$, we assume that the machine starts at time $t = 0$ at position $\theta = 0$, then the resulting motion is the sum of a deterministic exponential and correlated random fluctuation terms. The load eventually settles into correlated random motion of standard deviation $\sigma/\sqrt{2\rho}$ around the machine dwell conformation $\mu$. The effect of the load drag and stiffness is to delay the transitions by "rounding off" the instantaneous step transition in $\mu(t)$ with step time constant $\rho^{-1}$ s. Therefore, to be effective, a step-smoothing algorithm must take into account this delayed transition.

This is a continuous-time stochastic process, but the experimental angular measurements are available at the sampling interval $\Delta t$. Therefore, we need to find a discrete-time version of the model. The simple *Euler method* obtains:

$$\theta_{t+1} = \theta_t + \frac{\kappa}{\xi}(\mu - \theta_t)\Delta t + \sqrt{\frac{2k_B T \Delta t}{\xi}}\varepsilon_t = \left(1 - \frac{\kappa \Delta t}{\xi}\right)\theta_t + \frac{\kappa \Delta t}{\xi}\mu + \sqrt{\frac{2k_B T \Delta t}{\xi}}\varepsilon_t \qquad (17)$$

where $\theta_t = \theta(t\Delta t)$ for the time index $t = 0, 1 \ldots T$ with $\theta(0) = 0$ (we note that although there are a range of generally more accurate methods for discretising such SDEs, most are no more accurate for this particular model and so in this context there is no particular advantage to using a higher order integration scheme). This is also a discrete-time, first order autoregressive (AR) model in the form:

$$\theta_{t+1} = a_1 \theta_t + (1 - a_1)\mu_t + \varepsilon'_t \qquad (18)$$

where $\varepsilon'_t$ is an independent, zero-mean, constant variance Gaussian process. Therefore, this model is a special case of the implicit model in algorithm L1-PWC-ARP with ($P = 1$) described above, and we can estimate the quantity $a_1 = 1 - \kappa \Delta t / \xi$ directly from experimental time series using the autocorrelation at one time lag $\Delta t$ of measured bead time traces $\theta_t$.

In the main text we explore simulations of the rotary bacterial flagellar motor with typical experimental parameters based on calculated bead load particle of diameter 0.15μm, $\xi = 0.01 k_B T$ s, and measured bacterial flagellar properties[7] $\kappa = 100 k_B T / \text{rad}$. This gives a smooth step time constant of $\tau = 10^{-4}$ s. At $5\tau$ after an instantaneous transition in the equilibrium angle, the bead will have settled to within 1% of the steady state $\mu$. With a sampling rate of $\Delta t = 1/104448\ s$ (chosen to match one of our high-resolution experimental recording setups, although values that match any particular experiment can be used here), the discrete-time first-order AR1 coefficient is $a_1 = 1 - 0.096 = 0.904$, and the standard deviation of the noise term $\varepsilon_t$ is $\sqrt{2\Delta t/0.01} = 0.044$.

**Step-smoothing and bump-hunting algorithm performance comparisons.** Figure 3 (main text) describes nine simulated test cases produced by varying: the symmetry of the discrete state locations (that is, by randomly displacing the state locations from equal spacing), the distribution of dwell times (by changing the gamma shape parameter *k*), the dominant symmetry (e.g. the number of discrete states), the average speed of rotation (that is, the number of revolutions per second, controlled by scaling the dwell times), and the stiffness parameter $\kappa$.

Figure 2 (main text) shows the typical output from the discrete-time model, and Figure 3 (main text) shows the performance of a range of step-smoothing algorithms applied to this test data. We test L1-PWC, L1-PWC-AR1, median filtering[8], the Chung-Kennedy filter[9], and the Kalafut-Visscher step-finding methods[10]. We compare the performance of these methods in terms of the accuracy of their ability to extract the unobserved motor position $\mu(t)$ using the *mean absolute error*:

$$MAE = \frac{1}{T}\sum_{t=1}^{T}|\mu_t - m_t| \qquad (19)$$

and smaller is better. Also, the *relative absolute roughness*:

$$RAR = \sum_{t=1}^{T-1}|m_{t+1} - m_t| / \sum_{t=1}^{T-1}|\mu_{t+1} - \mu_t| \qquad (20)$$

identifies over- and under-smoothing relative to the known, motor position time series, the closer to unity the better. Note that if the *MAE* = 0, then the *RAR* = 1 (although *RAR* = 1 does not necessarily give *MAE* = 0, thus, it is important to interpret the performance with respect to *both* quantities).

Step-smoothing algorithm parameters are optimized on this test data to achieve the best MAE and RAR values. For the L1-PWC algorithm the optimal parameter values were $\gamma = 50$, and for L1-PWC-AR1, $\gamma = 1$, $P = 1$ and $a_1 = 1 - \kappa \Delta t / \xi$. For the median filter, the only parameter is the window size, and as expected the optimum size was found to be the average dwell time. For the Chung-Kennedy filter, the maximum size of all forward/backward moving average predictors plays a similar role to the window size in the median filter. In our implementation, we included predictors of all window sizes up to this maximum window size, and extensive experimentation found that setting this to half the average dwell time optimized performance. Because of the non-instantaneous stepping of the Langevin dynamics, for the nonlinearity $p > 0$, we found that this algorithm introduced numerous spurious steps and non-smoothness that degraded the performance considerably. Therefore, we found that having no nonlinearity (i.e setting $p = 0$) led to the best performance overall, because it was the smoothest possible filter and so was able to perform well for the longer dwell times. The Kalafut-Visscher filter has no explicitly tunable parameters, although we have found that the results depend heavily on the length of the time series.

Bump-hunting algorithm comparisons were made in terms of the median and interquartile range (25% – 75% range) of the recovered number of discrete states (see Supplementary Tables 1 and 2). Algorithm parameters were optimized to achieve the best recovery performance. For the ECF-Bump algorithm, the analysis symmetries (frequencies) ranged from zero to 120 steps per revolution, and the nonlinear threshold was set to retain the top 20% largest square magnitude frequencies. The linear threshold was set at 80 steps per revolution. The histogram FFT algorithm used 128 histogram bins and 128-point FFT. The kernel density peak-picking algorithm had a Gaussian kernel with bandwidth parameter of 0.02 rads$^2$, and peaks smaller than 10% of the maximum peak amplitude were discarded.

**Distribution fitting to dwell times.** Standard maximum likelihood techniques minimizing the negative log-likelihood have been used to fit each distribution model to the dwell times obtained from bacterial flagellar motor and F$_1$-ATPase time-angle traces (see below for explicit details of the double exponential model). For distribution model comparison, the Bayesian Information Criterion is calculated as[11]:

$$BIC = -2L + p \log N \quad (21)$$

where $p$ is the number of free parameters in each distribution model, $N$ is the number of dwell times, and $L$ is the log-likelihood of the distribution model. For the exponential, $p = 1$, gamma, lognormal and double exponential, $p = 2$, and for the generalized Pareto, $p = 3$. For the gamma with fixed $k$, $p = 1$.

When there are multiple subsets of dwell times that require separate distribution models per dwell state, the total *BIC* is obtained by adding the *BIC* for each separate model – this is consistent with assumption that each dwell state is independent of the others.

**L1-PWC-AR1 autoregressive parameter estimation.** To use the L1-PWC-AR1 algorithm, we use the standard covariance method for autocorrelation analysis to estimate the parameter $a_1$, which is the autocorrelation at a time lag of one sample. This requires manual identification of a sufficiently long section of the signal where the molecular machine is stationary. In the real F$_1$-ATPas data we studied, unambiguous, long dwells are frequent so that this approach is straightforward.

**Double exponential distribution.** For more than one reaction cascaded together, a more complex process than the simple Poisson process is usually a better model for the observed discrete state dwell times. Assuming that one reaction has to wait for the other to finish, the total dwell time will be a random variable that is the sum of two exponentially-distributed dwell times $T = T_1 + T_2$ with rate parameters $k_1$, $k_2$. Then the distribution of $T$ is the convolution of the distribution of $T_1$ and $T_2$. This becomes the product of the moment generating functions of the individual distributions:

$$M_T(s) = \frac{k_1}{s+k_1} \frac{k_2}{s+k_2} \quad (22)$$

Inverting the moment generating function gives the distribution:

$$p(T) = \frac{k_1 k_2}{k_2 - k_1} \left[ \exp(-k_1 T) - \exp(-k_2 T) \right] \quad (23)$$

with mean $(k_1 + k_2)/(k_1 k_2)$ and variance $(k_1^2 + k_2^2)/(k_1^2 k_2^2)$.

To fit the rate parameters given a set of dwell times $T_i$, $i = 1, 2 \ldots N$, we can maximize the likelihood, which is equivalent to minimizing the negative log-likelihood:

$$\hat{k}_1, \hat{k}_2 = \underset{k_1, k_2}{\arg\min} \left[ -N \log \frac{k_1 k_2}{k_2 - k_1} - \sum_{i=1}^{N} \log\left(\exp(-k_1 T_i) - \exp(-k_2 T_i)\right) \right] \quad (24)$$

This can be solved using a variety of generic nonlinear optimization techniques, with the constraint $k_1, k_2 > 0$. Confidence intervals for the rate parameters are obtained by 1000 bootstrap resampling operations[11].

The degenerate case where $k_1 = k_2$ is the gamma distribution with scale parameter $k = 2$.

|  | ECF-Bump | Kernel density with peak finding |
|---|---|---|
| *Default* (26) | 26 (0.0) | 27 (1.5) |
| *20% dwell aperiodicity* (26) | 26 (0.5) | 27 (1.5) |
| *Gamma dwell times, k* = 2 (26) | 26 (0.0) | 29 (1.3) |
| *Gamma dwell times, k* = 10 (26) | 26 (0.0) | 29 (2.5) |
| 30 *dwell locations* (30) | 30 (0.0) | 32 (3.3) |
| 40 *dwell locations* (40) | 40 (0.0) | 46 (5.0) |
| 50 *revs/sec* (26) | 26 (0.0) | 39 (5.3) |
| 100 *revs/sec* (26) | 26 (0.0) | 45 (3.8) |
| *Flagellar stiffness, κ* = 50 (26) | 26 (0.0) | 47 (8.8) |

*Supplementary Table 1: Bump-hunting method symmetry detection performance under diverse conditions.* Ability of two different bump-hunting methods to recover the dominant symmetry in the distribution of states for the nine test cases of simulated bacterial flagellar motor rotation time series described in Figure 3 (main text). The figures are the median dominant symmetry over five replications, and the associated interquartile range (difference between 25th – 75th percentile) in brackets. The bracketed number in the first column is the true symmetry. Algorithm ECF-Bump is described in Supplementary Methods. Kernel density with peak finding estimates the distribution of discrete states using the kernel density method, then counts the number of peaks in the estimated distribution.

| | Average exponential dwell time | | | | |
|---|---|---|---|---|---|
| Number of dwell locations | 2.5ms | 1.5ms | 1.0ms | 0.75ms | 0.50ms |
| *ECF-Bump* | | | | | |
| 20 | 20.0 (0.0) | 20.0 (0.0) | 20.0 (0.0) | 20.0 (0.0) | 20.0 (0.0) |
| 30 | 30.0 (0.0) | 30.0 (0.0) | 30.0 (0.0) | 30.0 (0.0) | 30.0 (0.0) |
| 40 | 40.0 (0.0) | 40.0 (0.0) | 40.0 (0.0) | 40.0 (0.0) | 40.0 (0.0) |
| 50 | 50.0 (0.0) | 50.0 (0.0) | 50.0 (2.5) | 50.0 (0.0) | 54.0 (8.0) |
| 60 | 60.0 (0.0) | 60.0 (0.0) | 60.0 (4.0) | 60.0 (0.0) | (no result) |
| *Histogram with FFT* [12] (see caption) | | | | | |
| 20 | 19.0 (0.0) | 19.0 (1.0) | 20.0 (1.0) | 20.0 (0.0) | 20.0 (0.0) |
| 30 | 29.0 (0.0) | 29.0 (0.0) | 29.0 (0.0) | 29.0 (0.0) | 30.0 (1.0) |
| 40 | 39.0 (0.0) | 39.0 (0.0) | 39.0 (34.0) | 8.0 (16.0) | 14.5 (13.0) |
| 50 | 5.5 (45.0) | 4.5 (9.0) | 4.0 (14.0) | 15.5 (15.0) | 15.5 (24.0) |
| 60 | 20.5 (47.0) | 13.0 (16.0) | 7.5 (9.0) | 14.0 (17.0) | 14.0 (18.0) |
| *Kernel density with peak finding* (see caption) | | | | | |
| 20 | 21.0 (0.0) | 21.0 (1.0) | 21.0 (1.0) | 21.0 (1.0) | 21.0 (1.0) |
| 30 | 30.5 (1.0) | 30.0 (1.0) | 31.0 (2.0) | 32.5 (3.0) | 34.0 (2.0) |
| 40 | 37.5 (2.0) | 38.0 (3.0) | 38.0 (2.0) | 37.0 (1.0) | 39.0 (2.0) |
| 50 | 40.5 (3.0) | 37.0 (2.0) | 37.0 (2.0) | 36.0 (4.0) | 38.0 (2.0) |
| 60 | 40.0 (3.0) | 37.0 (4.0) | 37.0 (4.0) | 37.5 (1.0) | 36.0 (4.0) |

*Supplementary Table 2: Bump-hunting method symmetry detection at different stepping rates.* Performance of different bump-hunting methods at finding the dominant symmetry in the distribution of states of simulated bacterial flagellar motor rotation time series with exponential dwell times, over a wide range of dominant symmetries. Each entry shows the median estimated state periodicity, with the interquartile range (25th – 75th percentile) in brackets. Algorithm ECF-Bump is described in Supplementary Methods. No result indicates that no dominant peak in the ECF could be found. The histogram with FFT method first estimates the distribution of states using a histogram, then finds the fast Fourier transform of that histogram; the largest peak in the spectrum is the estimated periodicity (method used in Sowa *et al.* 2005). Kernel density with peak finding estimates the distribution of states using the kernel density method, then counts the number of peaks in the estimated distribution.

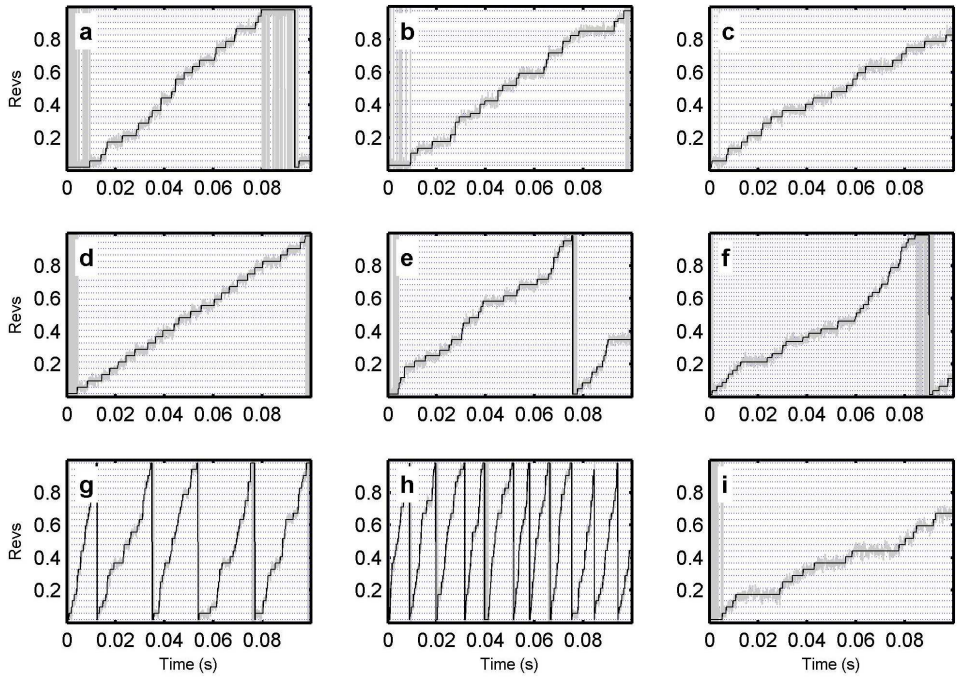

*Supplementary Figure 1: Simulations of bacterial flagellar motor time series under diverse conditions.* Nine test cases of simulated bacterial flagellar motor time traces; gray line is measured bead angular position $\theta_t$, black line the (unobservable) motor position $\mu_t$. Blue horizontal lines are the discrete state locations. (a) Default case: 26 regularly spaced states, exponential dwell times, flagellar hook stiffness $\kappa = 100 k_B T/$ rad, 10 revs/sec. (b) As default, but with 20% dwell location asymmetry (see Supplementary Methods). (c) With gamma-distributed dwell times, $k = 2$. (d) Gamma dwell times, $k = 10$. (e) 30 states. (f) 40 states. (g) 50 revs/sec. (h) 100 revs/sec. (i) Flagellar hook stiffness $\kappa = 50\ k_B T/$rad.

**Supplementary References**

1. Donoho, D.L. For most large underdetermined systems of equations, the minimal l(1)-norm near-solution approximates the sparsest near-solution. *Communications on Pure and Applied Mathematics* **59**, 907-934 (2006).
2. Strong, D. & Chan, T. Edge-preserving and scale-dependent properties of total variation regularization. *Inverse Problems* **19**, S165-S187 (2003).
3. Boyd, S.P. & Vandenberghe, L. *Convex optimization*, xiii, 716 p. (Cambridge University Press, Cambridge, UK ; New York, 2004).
4. Kim, S.J., Koh, K., Boyd, S. & Gorinevsky, D. L1 Trend Filtering. *SIAM Review* **51**, 339-360 (2009).
5. Becker, O.M. *Computational biochemistry and biophysics*, xii, 512 p. (M. Dekker, New York, 2001).
6. Candes, E.J. Modern statistical estimation via oracle inequalities. *Acta Numerica* **15**, 257-326 (2006).
7. Block, S.M., Blair, D.F. & Berg, H.C. Compliance of Bacterial Flagella Measured with Optical Tweezers. *Nature* **338**, 514-518 (1989).
8. Arce, G.R. *Nonlinear signal processing: a statistical approach*, xx, 459 p. (Wiley-Interscience, Hoboken, N.J., 2005).
9. Chung, S.H. & Kennedy, R.A. Forward-Backward Nonlinear Filtering Technique for Extracting Small Biological Signals from Noise. *Journal of Neuroscience Methods* **40**, 71-86 (1991).
10. Kalafut, B. & Visscher, K. An objective, model-independent method for detection of non-uniform steps in noisy signals. *Computer Physics Communications* **179**, 716-723 (2008).
11. Hastie, T., Tibshirani, R. & Friedman, J.H. *The elements of statistical learning : data mining, inference, and prediction : with 200 full-color illustrations*, xvi, 533 p. (Springer, New York, 2001).
12. Sowa, Y. et al. Direct observation of steps in rotation of the bacterial flagellar motor. *Nature* **437**, 916-919 (2005).